# Development of 15kA/cm² Fabrication Process for Superconducting Integrated Digital Circuits

Liliang Ying, Xue Zhang, Guixiang He, Weifeng Shi, Hui Xie, Linxian Ma, Hui Zhang, Jie Ren, Wei Peng, and Zhen Wang

*Abstract* — A new fabrication process for superconducting integrated digital circuits is reported. We have developed the "SIMIT Nb04" fabrication technique for superconducting integrated circuits with Nb-based Josephson junctions based on the validated "SIMIT Nb03" process and Chemical Mechanical Planarization (CMP) technology. Seven Nb superconducting layers and one Mo resistor layer are included in the "SIMIT Nb04" process with 19 mask levels. The device structure is composed of active layers including junctions at the bottom, two passive transmission line (PTL) layers in the middle and a DC power layer at the top. The circuit fabrication started with the fabrication of Mo resistors with a target sheet resistance $R_{sh}$ of 3 Ω, followed by the deposition of Nb/Al-AlO$_x$/Nb trilayer Josephson-junction with a target critical current density $J_c$ at 15 kA/cm². To increase the Al-AlO$_x$ barrier layer etching's repeatability, an additional barrier protection layer was applied. To accomplish high-quality planarization, we created a planarization procedure coupled with dummy filling. To assess the process dependability and controllability, a set of process control monitors (PCMs) for monitoring fabrication and design parameters was designed and monitored. The successful manufacturing and testing of a few small-scale circuits, like our standard library cells, further attests to the viability of our fabrication process for superconducting integrated circuits.

*Index Terms*—Josephson device fabrication; SFQ circuits; Superconducting LSI; Superconducting electronics fabrication.

## I. INTRODUCTION

Rapid single flux quantum (RSFQ) circuits have the advantages of high operating frequencies and lower power consumptions. Numerous efforts have been made to develop RSFQ large-scale integrated (LSI) circuit fabrication technologies mainly based on Nb/Al-AlO$_x$/Nb Josephson junctions (JJs). For example, MIT Lincoln Laboratory [1]-[3] and HYPRES [4]-[6] in the USA, National Institute of Advanced Industrial Science and Technology in Japan [7]-[9], Leibniz Institute of Photonic Technology in Germany [10] have reported such circuits.

Using our usual fabrication method, we have been manufacturing several superconducting single flux quantum (SFQ) circuits (SIMIT Nb03 [11][12]). In this paper, we report on the development of our new fabrication process (SIMIT Nb04). Seven Nb superconducting layers and one Mo resistor layer are included. The process started with the fabrication of Mo resistors with a target sheet resistance $R_{sh}$ of 3 Ω, followed by the deposition of Nb/Al-AlO$_x$/Nb trilayer Josephson-junction with a target critical current density $J_c$ at 15 kA/cm². The wafer process flow is briefly discussed. Successful fabrication and testing of a few small-scale circuits are demonstrated.

## II. FABRICATION PROCESS

### A. Overview

The target parameters of the layers and the minimum feature sizes are summarized in Table I. The process includes a resistor layer RN0, a trilayer MN0/AN0/JN0, a wiring MP1, two passive transmission line (PTL) layers in the middle, two ground plane MP2 and MP5, a DC power layer at the top, a contact pad metallization PP2. Nb/Al-AlO$_x$/Nb trilayers were fabricated in a multi-chamber dc magnetron sputtering system, in which Nb and Al films were deposited in two separated vacuum chambers, and the load-lock chamber was used for aluminum oxidation. The thickness of AlO$_x$ was controlled by exposure time and the O$_2$ partial pressure, which corresponds to the target critical current density $J_c$ for 15 kA/cm². Mo resistors with the target sheet resistance $R_{sh}$ of 3 Ω was deposited by DC magnetron sputtering. SiO$_2$ interlayer insulators are deposited with an plasma-enhanced chemical vapor deposition (PECVD). The patterning is achieved by using a new i-line stepper Canon FPA-3000 EX4 with 248 nm exposure wavelength. Nb film is etched by Inductively Coupled Plasma (ICP) etching system with endpoint detector, while the SiO$_2$ insulator is etched by Reactive Ion Etching system (RIE).

For the junction barrier Al-AlO$_x$ layer etching, most of the fabrication processes uses ion beam etching (IBE) [9] or wet etching [14][15]. However, sidewalls formed in the IBE process may create additional conductive paths in the insulation layer t leading to a current leakage in the junctions. Moreover the automatic wet etching used in standard process (Nb03) can avoid this problem, but it may produce some black

This work was supported by the Strategic Priority Research Program of Chinese Academy of Sciences (Grant No. XDA18000000), Shanghai Science and Technology Committee (Grant No. 21DZ1101000), Chinese Academy of Sciences Key Technology Talent Program, the National Natural Science Foundation of China under Grant No. 62171437 and No. 92164101, and the National Key R&D Program of China under Grant No. 2021YFB0300400. (*Corresponding author: Liliang Ying, Jie Ren*)
The authors are with Shanghai Institute of Microsystem and Information Technology, Chinese Academy of Sciences, Shanghai 200050, China and also with CAS Center for Excellence in Superconducting Electronics (e-mail: llying@mail.sim.ac.cn, jieren@mail.sim.ac.cn).
Liliang Ying, Jie Ren, Wei Peng and Zhen Wang are also with University of Chinese Academy of Science, Beijing 100049, China.

reaction products. To solve these problems, we used a very thin SiO$_2$ film as a barrier protective layer and optimized IBE etching parameters to improve the reproducibility of Al-AlOx barrier layer etching.

TABLE I
LAYER PARAMETER AND MASK INFORMATION

| Layer | Material | Description | Thickness (nm) | Min. size (μm) | Min. space (μm) |
|---|---|---|---|---|---|
| RN0 | Mo | Resistor | 30 ± 5 | 1.6 | 1.0 |
| CN0 | SiO$_2$ | Insulation | 100 ± 10 | 1.0 | 1.0 |
| MN0 | Nb | Base electrode | 150 ± 5 | 1.0 | 1.0 |
| AN0 | Al-AlO$_x$ | Junction barrier | 12 ± 0.5 | 1.5 | 2.0 |
| JN0 | Nb | Counter electrode | 150 ± 5 | 0.5 | 1.0 |
| IN0J | SiO$_2$ | Insulation | 70 ± 10 | 0.7 | 1.0 |
| IN0M | SiO$_2$ | Insulation | 220 ± 10 | 1.0 | 1.0 |
| MP1 | Nb | Superconductor | 300 ± 15 | 1.0 | 0.8 |
| IP1 | SiO$_2$ | Insulation | 120 ± 10 | 1.0 | 1.0 |
| MP2 | Nb | Superconductor | 200 ± 10 | 1.0 | 0.8 |
| IP2 | SiO$_2$ | Insulation | 120 ± 10 | 1.0 | 1.0 |
| MP3 | Nb | Superconductor | 200 ± 10 | 1.0 | 0.8 |
| IP3 | SiO$_2$ | Insulation | 120 ± 10 | 1.0 | 1.0 |
| MP4 | Nb | Superconductor | 200 ± 10 | 1.0 | 0.8 |
| IP4 | SiO$_2$ | Insulation | 120 ± 10 | 1.0 | 1.0 |
| MP5 | Nb | Superconductor | 200 ± 10 | 1.0 | 0.8 |
| IP5 | SiO$_2$ | Insulation | 120 ± 10 | 1.0 | 1.0 |
| MP6 | Nb | Superconductor | 200 ± 10 | 1.0 | 0.8 |
| PP2 | Ti/Au | Contact pad | 108 ± 5 | 3.0 | 3.0 |

*B. Barrier protective layer process*

An additional barrier protective layer was carried out to improve the reproducibility of Al-AlOx barrier layer etching and the reliability of the junctions, as shown in Fig. 1. Following the inductively coupled plasma RIE (ICP-RIE) formation of the Nb/Al-AlO$_x$/Nb trilayer, a 10-nm SiO$_2$ additional barrier protective layer (SiO$_2$) was applied using low-power PECVD. Following the optimized IBE, the AlO$_x$ tunnel barrier (AN0) was patterned and then etched by RIE with CHF$_3$ at 2 Pa. After the trilayer patterning, a 600-nm SiO$_2$ was deposited using PECVD, then the wafer was planarized by CMP. Finally vias in IN0J and IN0M were separately made by RIE.

*C. Circuit fabrication*

After 30-nm Mo (RN0) was first fabricated onto 100mm diam Si wafer with 300nm-thick thermal oxidation SiO$_2$ on the surface by dc magnetron sputtering and reactive ion etching (RIE), a 100-nm SiO$_2$ film (CN0) was deposited by PECVD, and etched using RIE with CHF$_3$ at 2 Pa. Then Nb/Al-AlO$_x$/Nb trilayers was deposited. First, an 150-nm-thick Nb film (MN0) was deposited in 0.7Pa Ar gas atmospheres and 2.0 A sputtering current with water cooling the substrate during sputtering. Then a 12-nm-thick Al layer was deposited at $P_{Ar}$ = 0.5 Pa and a sputtering current of 0.5 A. The Al film surface wad oxidized by Ar and 15%-O$_2$ mixture gas in load-lock chamber at pressure of 4 Pa for 10 to 20 mins. The counter Nb layer with 150nm thickness was deposited on the AlO$_x$ to finish junction. The counter Nb film was etched by inductively coupled plasma RIE (ICP-RIE). Then a 10-nm SiO$_2$ additional barrier protective layer (SiO$_2$) was deposited using low-power PECVD and tched by RIE with CHF$_3$ at 2 Pa. The optimized IBE was used to pattern the AlO$_x$ tunnel barrier (AN0).

After 600-nm SiO$_2$ deposited and the planarized, vias in IN0J and IN0M layers were separately etched by RIE. The Nb wiring (MP1) with 300-nm thickness was deposited and patterned. Then another 600-nm SiO$_2$ layer (IP1) was deposited, and vias in IP1 were fabricated with the same parameters after the planarization. A 200-nm Nb ground plane (MP2) was deposited and patterned. IP2-MP6 was formed using the same processing conditions as IP1 and MP2. Fig. 2 shows a scanning electron microscope image of a cross-section of the fabricated circuit.

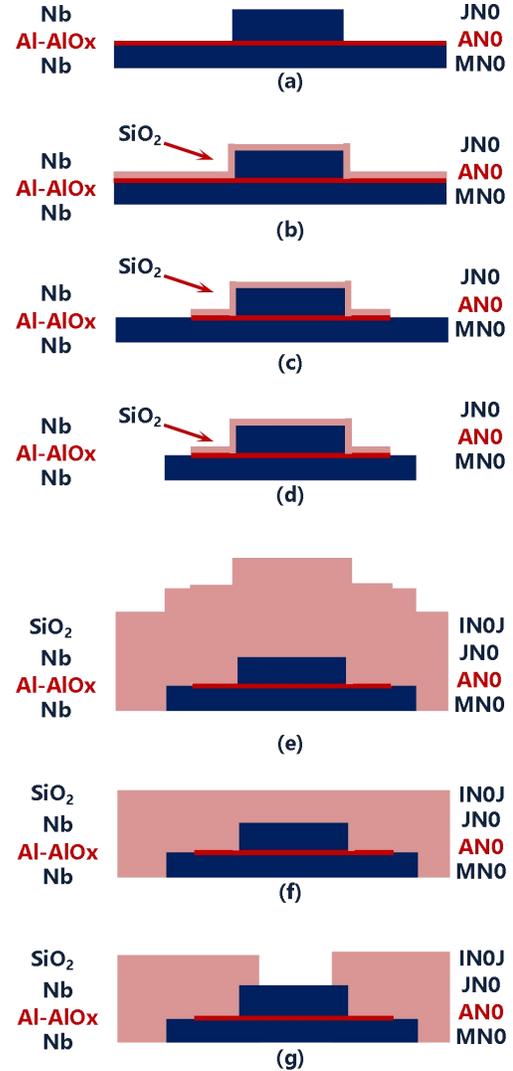

Fig. 1. Schematic illustration of the process steps for a Josephson junction. (a) counter electrodes Nb etching (JN0) ; (b) a thin SiO$_2$ film as an additional barrier protective layer was deposited using PECVD; (c) the AlOx tunnel barrier (AN0) was patterned, than etched by RIE and IBE; (d) base electrode Nb etching (MN0); (e) SiO$_2$ deposition using PECVD; (f) planarization by CMP; (e) SiO$_2$ etching for vias in IN0J.

During the wafer fabrication process, the process parameters such as lithography resolutions and alignment shifts, etched depths, and actual film thicknesses have been monitored using the on-wafer process control monitor (PCM) patterns. Details of the PCM fabrication technology evaluations are presented elsewhere [16].

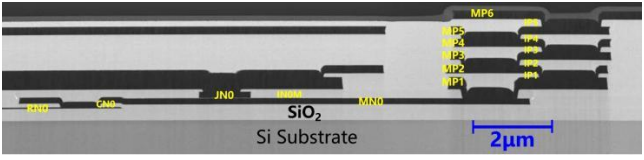

Fig. 2. Cross-section of a device fabricated by SIMIT standard fabrication process.

### III. TEST RESULTS

The fabrication process has been evaluated by electrical testing using PCM circuits [16]. The probability of the flaws as well as circuit parameters like $J_c$ and $R_{sh}$ was selected as evaluation criteria. It has become a standard to provide feedback on the PCM results to advance the fabrication technology.

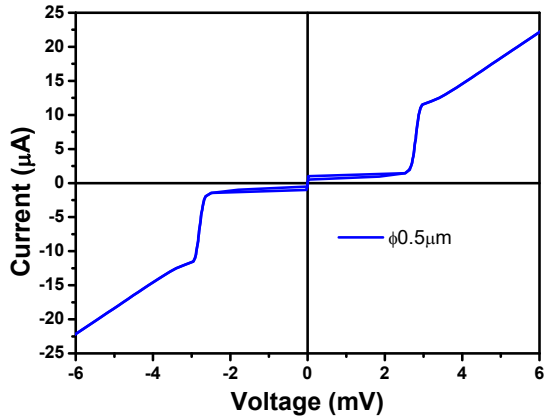

Fig. 3. Typical I-V characteristics for Nb/Al-AlO$_x$/Nb junctions fabricated by our standard process.

By submerging chips in liquid helium at 4.2 K, the current-voltage (I-V) of circular junctions of various diameters was measured using a four-wire measurement method. Fig. 3 shows typical I-V characteristics for Nb/Al-AlO$_x$/Nb junction. Parameters related to the junction property were extracted from current-voltage curves (I-V curves) of un-shunted JJs.

Josephson junctions' critical current density has an exponential relationship with those conditions as shown in Fig. 4. $J_c$ can be empirically fitted to $J_c \propto E^{-\alpha}$ [17], where $E$ denotes oxidation conditions (product of oxidation pressure and oxidation time) of Al layer in Nb/Al-AlO$_x$/Nb. By fitting experimental data on empirical formula from the process line, an accurate result of designed critical current density can be easily gotten, which is a useful guide for following SFQ device development and fabrication process. It should be noticed that the oxidation curves of different oxidation layers are distinct and partial pressure of oxygen also have a different exponential relationship at high $J_c$ or low $J_c$ region. Utilizing pure O$_2$ for the oxidation process, the resulting α = 0.46 in the low-$J_c$ region changes to 1.5 in the high-$J_c$ zone using oxidations in 15%O$_2$-Ar mixtures. The α value in the high-$J_c$ region is lower than our earlier reported data [11]. The reason for this is that to make the fitted curve more realistic, we have incorporated a lot more experimental data.

Small-scale circuits were designed and fabricated using our newly 15-kA/cm$^2$ Nbjunction technology SIMIT Nb04. OCTOPUX, an automated system specifically for evaluating superconducting circuits, was used to test the circuits at low frequencies. The margin comparison of the same circuit fabricated using various processes is shown in Fig. 5. The typical experimental DC bias margins were in the range of -40% to +40% for most of the gates. The same cell using different processes shows similar bias margins. This clearly illustrates the suitability of the novel technique for superconducting integrated digital circuits; it may enhance circuit integration without lowering circuit quality.

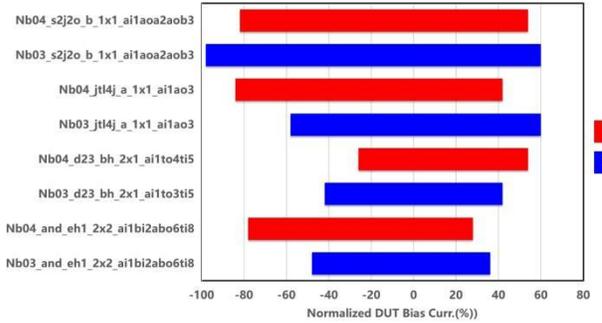

Fig. 5. Margin comparison of the same circuit fabricated by different process.

### IV. CONCLUSION

Based on the validated "SIMIT Nb03" method and CMP technology, we have developed a new 15 kA/cm$^2$ fabrication technique "SIMIT Nb04" featuring Nb/Al-AlOx/Nb Josephson junctions for superconducting integrated circuits. The critical current density $J_c$ and minimum diameter of the junction achieved are 15 kA/cm$^2$ and 0.5 μm, respectively. The fabrication parameters for the "SIMIT Nb04" process have been systematically optimized. To increase the consistency of Al-AlO$_x$ barrier layer etching and the reliability of sub-micron connections, a second barrier protection layer was used. We investigated the dependence of $J_c$ on O$_2$ exposure $E$ for high current density $J_c$, whose values were determined with PCM junctions. Several small-scale circuits have been successfully constructed and tested at low frequencies, like RSFQ logic gates. The typical experimental operating margins for the bias current were found to be similar to those from "SIMIT Nb03" process. As a result, it is demonstrated that the superconducting integrated digital circuits can be produced using our innovative technique, "SIMIT Nb04."

### ACKNOWLEDGMENT

The authors would like to thank Masaaki Maezawa for the valuable discussion. The fabrications were performed in the Superconducting Electronics Facility (SELF) of Shanghai Institute of Microsystem and Information Technology.

## REFERENCE

[1] S. K. Tolpygo, E. B. Golden, T. J. Weir, V. Bolkhovsky, "Increasing integration scale of superconductor electronics beyond one million Josephson junctions", *J. Phys.: Conf. Ser.,* **vol. 1559**, 012002, 2020;

[2] S. K. Tolpygo, V. Bolkhovsky, T. J. Weir, et al., "Inductance of Circuit Structures for MIT LL Superconductor Electronics Fabrication Process With 8 Niobium Layers", *IEEE Trans. Appl. Supercond.*, **vol. 25, No. 3**, 1100905, June 2015;

[3] S. K. Tolpygo, E. B. Golden, T. J. Weir, and V. Bolkhovsky, " Inductance of superconductor integrated circuit features with sizes down to 120 nm", *Supercond. Sci. Technol.*, **vol. 34,** 085005, June 2021;

[4] S.K. Tolpygo, D. Yohannes, R.T. Hunt et al., "Analysis of Multilayer Devices for Superconducting Electronics by High-Resolution Scanning Transmission Electron Microscopy and Energy Dispersive Spectroscopy", *IEEE Trans. Appl. Supercond.*, **vol. 17, No. 2**, 946-951, June 2007;

[5] S. K. Tolpygo, D. Amparo, D. T. Yohannes,et al., "Process-Induced Variability of Junctions in Superconductor IntegratedCircuits and Protection Against It", *IEEE Trans. Appl. Supercond.,* **Vol. 19, No. 3**, 135, June 2009;

[6] S. K. Tolpygo, D. Amparo, "Fabrication process development for superconducting VLSI circuits: minimizing plasma charging damage", *J. Phys.: Conf. Ser.,* **vol. 97**, 01227, 2008;

[7] N. Takeuchi, S. Nagasawa, F. China.et al., "Adiabatic quantum-flux-parametron cell library designed using a 10 kA cm$^{−2}$ niobium fabrication process", *Supercond. Sci. Technol.*, **vol. 30,** 035002, January 2017;

[8] T. Satoh, K. Hinode, H. Akaike, et al., " Fabrication process of planarized multi-layer Nb integrated circuits", *IEEE Trans. Appl. Supercond.*, **vol. 15, No. 2**, 78-88, June 2005;

[9] M. Hidaka, S. Nagasawa, K. Hinode, and T. Satoh, "Device Yield in Nb-Nine-Layer Circuit Fabrication Process", *IEEE Trans. Appl. Supercond.*, **vol. 23, No. 3**, 1100906, June 2013;

[10] Rapid Single Flux Quantum (RSFQ) – Design Rules for Nb/Al$_2$O$_3$-Al/Nb-Process at Leibniz IPHT, Version 10.03.2017: RSFQ1H-1.6, available online: https://www.fluxonics.de/fluxonics-foundry/;

[11] L. Ying, X. Zhang, M. Niu, J. Ren, W. Peng, M. Maezawa, and Z. Wang, "Development of Multi-Layer Fabrication Process for SFQ Large Scale Integrated Digital Circuits", *IEEE Trans. Appl. Supercond.*, **vol. 31, No. 5,** 1301504, August 2021;

[12] X. Gao, Q. Qiao, M. Wang, M. Niu, H. Liu, M. Maezawa, J. Ren, and Z. Wang, "Design and Verification of SFQ Cell Library for Superconducting LSI Digital Circuits", *IEEE Trans. Appl. Supercond.*, **vol. 31, No. 5,** 1101105, August 2021;

[13] Y. Wu, L. Ying, G. Li, et al., "Film Stress Influence on Nb/Al-AlOx/Nb Josephson Junctions", *IEEE Trans. Appl. Supercond.*, **vol. 29, No. 5**, 1102105, August 2019;

[14] R. Monaco, R. Cristiano, L. Frunzio, and C. Nappi, "Investigation of low-temperature I-V curves of high -quality Nb/AI-AlOx/Nb Josephson junctions", *J. Appl. Phys.,* **71 (4),** 1888-1892, 1991;

[15] M. Maezawa, M. Ochiai, H. Kimura, F. Hirayama, and M. Suzuki, "Design and Operation of RSFQ Cell Library Fabricated by Using 10-kA/cm$^2$ Nb Technology", *IEEE Trans. Appl. Supercond.*, **vol. 17, No. 2**, pp. 500 - 504, June 2007;

[16] X. Zhang, L. Ying, et. al., "Process Control Monitoring for Fabrication Technology of Superconducting Integrated Circuits", *IEEE Trans. Appl. Supercond.*, **vol. 31, No. 5,** 1101206, August 2021;

[17] A.W. Kleinsasser, R.E. Miller, W.H. Mallison, "Dependence of Critical Current Density on Oxygen Exposure in Nb-Al-Nb Tunnel Junctions", *IEEE Trans. Appl. Supercond.*, **5,** 26-30, 1995.